\documentclass[conference]{IEEEtran}
\IEEEoverridecommandlockouts
\usepackage{cite}
\usepackage{amsmath,amssymb,amsfonts}
\usepackage{algorithmic}
\usepackage{graphicx}
\usepackage{textcomp}
\usepackage{tabularx}
\usepackage{xcolor}
\def\BibTeX{{\rm B\kern-.05em{\sc i\kern-.025em b}\kern-.08em
    T\kern-.1667em\lower.7ex\hbox{E}\kern-.125emX}}
\begin{document}

\title{From Safety Documentation to Safety Knowledge Support: An Evidence-Grounded LLM Framework for Medical Devices\\
\thanks{This project has received funding from the European Union’s EU Framework Program for Research and Innovation Horizon Europe under Grant Agreement No. 101169197. Views and opinions
expressed are, however, those of the authors only and not necessarily reflect those of the European Union or Fraunhofer Gesellschaft or RPTU Kaiserslautern-Landau.}
}

\author{\IEEEauthorblockN{Tuhinangshu Gangopadhyay}
\IEEEauthorblockA{\textit{Fraunhofer IESE}\\
\textit{RPTU Kaiserslautern-Landau}\\
Kaiserslautern, Germany \\
tuhinangshu.gangopadhyay@iese-extern.fraunhofer.de
}
\and
\IEEEauthorblockN{Rasmus Adler}
\IEEEauthorblockA{\textit{Fraunhofer IESE}\\
Kaiserslautern, Germany \\
rasmus.adler@iese.fraunhofer.de}
\and
\IEEEauthorblockN{Peter Liggesmeyer}
\IEEEauthorblockA{\textit{Fraunhofer IESE}\\
\textit{RPTU Kaiserslautern-Landau}\\
Kaiserslautern, Germany \\
peter.liggesmeyer@iese.fraunhofer.de}
\and
\IEEEauthorblockN{\centerline{Jan Reich}}
\IEEEauthorblockA{\textit{Fraunhofer IESE}\\
Kaiserslautern, Germany \\
jan.reich@iese.fraunhofer.de}
}

\maketitle

\begin{abstract}

Medical devices are becoming more software-intensive, connected, and AI-enabled. Their development requires risk-management evidence aligned with ISO 14971 and, for software, IEC 62304. This evidence must be kept consistent across requirements, design decisions, software changes, verification results, complaints, and post-market data. These tasks are costly and depend on scarce safety and domain experts.

Large language models (LLMs) may reduce parts of this effort because medical-device safety work is highly document-based. However, current LLM-based safety-engineering studies often address isolated methods, rely on generic prompting or public examples, and provide limited support for source links, traceability, uncertainty handling, lifecycle updates, and recorded expert review. This limits their use in regulated medical-device development.

This paper argues that the central research problem is not safety-text generation, but source-linked safety-knowledge support. We propose an evidence-grounded framework that connects device artifacts, controlled knowledge storage and retrieval, method-specific generation of candidate safety items, critique and uncertainty checks, and recorded expert review. The framework prepares, links, checks, and updates candidate safety artifacts for expert decision-making. It does not decide whether a device is safe and does not provide regulatory approval. We also outline an evaluation strategy using non-public or newly built medical-device case studies and expert reference analyses to assess coverage, correctness, relevance, traceability, duplicate rate, unsupported claims, and review effort.

\end{abstract}


\section{Introduction}

Medical devices are increasingly software-intensive, connected, and AI-enabled. This development supports new functions in diagnosis, therapy, rehabilitation, hospital logistics, and surgery. At the same time, it increases the effort needed to show that a device is safe for patients, users, and its intended use environment. Manufacturers must provide risk-management evidence and safety documentation aligned with standards such as ISO 14971 for risk management \cite{b1} and IEC 62304 for medical software lifecycle processes \cite{b2}.

Safety analysis is expert-intensive. Engineers must identify hazards, hazardous situations, failure modes, causal factors, foreseeable misuse, possible harm, and suitable risk-control measures. They must also keep these artifacts aligned with requirements, architecture, software design, verification evidence, production data, post-production data, and device changes. This creates a major cost and time burden, especially for small and medium-sized manufacturers. An MDR survey by DIHK, MedicalMountains, and SPECTARIS reports certification costs as the most frequently named reason for discontinuing products in the EU, cited by 91\% of respondents \cite{b3}.

LLMs are a promising support technology for this setting because safety engineering is strongly document-based. Device descriptions, requirements, hazard logs, FMEA tables, risk-control records, safety cases, test reports, user manuals, incident reports, and standards contain information that can support risk-management work. LLMs can help safety engineers draft candidate hazards, failure modes, causal links, risk controls, safety requirements, review questions, and safety-case fragments.

However, in regulated medical-device development, such support cannot be treated as independent safety assessment. Generated outputs need source grounding, traceability, uncertainty indicators, and expert review. A candidate hazard without evidence links is difficult to assess. A candidate risk control without links to requirements and verification evidence is difficult to justify. A generated risk score without recorded assumptions may even mislead reviewers.

As reviewed in Section III, recent work shows that LLMs can support tasks such as Hazard Analysis and Risk Assessment (HARA), Failure Mode and Effects Analysis (FMEA), Fault Tree Analysis (FTA), System-Theoretic Process Analysis (STPA), safety-requirements generation, and assurance-case analysis. Yet these studies often address isolated tasks, rely on prompt engineering, use public case studies, or focus on non-medical domains. This limits their use for medical devices, where device-specific knowledge, confidentiality, audit trails, rare high-severity harms, lifecycle updates, and fair evaluation are central concerns.

This paper argues that LLM-supported medical-device safety analysis should be framed as source-linked safety-knowledge support. The goal is not to let an LLM decide whether a device is safe. The goal is to help experts prepare, link, check, review, and update candidate safety artifacts in a controlled process.

The paper makes four contributions:

\begin{enumerate}

\item It summarizes recent patterns in LLM-assisted safety engineering and classifies them by safety task and dominant LLM method.
\item It identifies gaps that limit current approaches in regulated medical-device development.
\item It proposes a framework for evidence-grounded, traceable, reviewable, and updatable safety-knowledge support.
\item It defines an evaluation strategy based on non-public or newly built medical-device case studies, expert reference analyses, and safety-specific metrics.

\end{enumerate}

The remainder of the paper is structured as follows. Section II explains why medical-device safety analysis is a demanding research context. Section III reviews patterns in LLM-assisted safety engineering. Section IV derives the research gap. Section V presents the proposed framework. Section VI outlines the evaluation strategy. Section VII concludes the paper.

\section{Medical-Device Safety Analysis as a Research Context}

Medical-device safety analysis is not a final check before market release. It is a lifecycle activity. ISO 14971 requires manufacturers to identify hazards and hazardous situations, estimate and evaluate risks, select and verify risk-control measures, assess residual risk, and use production and post-production information \cite{b1}. IEC 62304 adds software lifecycle activities, including requirements, architecture, implementation, verification, maintenance, and risk-related software changes \cite{b2}.

This creates a demanding traceability problem. Safety artifacts must be linked to device functions, requirements, software items, design decisions, verification evidence, risk controls, and residual-risk decisions \cite{b1, b2, b4, b5}. When a requirement changes, when software is updated, or when post-market feedback reveals a new issue, the manufacturer must assess which safety artifacts are affected and whether additional risk control or verification is needed.

Medical devices also differ from many other technical systems because harm can depend on the device, the user, the patient, and the clinical context together. Consider an infusion pump. A safety analysis must not only consider hardware faults and software defects. It must also consider wrong dose entry, wrong flow rate, delayed occlusion detection, misleading alarms, use under time pressure, patient-specific sensitivity, and interaction with clinical workflows. The same technical failure may have different safety meaning depending on the intended use and the patient context.

The task becomes harder as devices become connected and AI-enabled. Modern medical devices may combine sensors, embedded software, networked services, cloud components, robotic functions, and machine-learning models. AI-enabled functions add further concerns because behavior can depend on training data, operating conditions, changes in input data, model limits, and uncertain predictions. Reported recalls and post-market safety events for AI-enabled medical devices show that safety evidence must also be maintained after release \cite{b6}. The TruDi case illustrates this lifecycle concern: incorrect localization events and patient injuries were reported after AI-supported functionality had been introduced, while AI was not identified as the reported cause of the errors \cite{b7}. Such cases show that safety analysis must account for changing device behavior, clinical use, and field feedback.

These properties make medical-device safety analysis a strong but high-risk use case for LLM support. LLMs may help draft candidate hazards, causes, risk controls, safety requirements, and review questions. Yet for regulated medical-device work, useful support requires source evidence, weak-support checks, qualified expert review, and lifecycle update support. The next section reviews whether current LLM-assisted safety-engineering work provides these capabilities.

\section{Structured Review of LLM-Assisted Safety Engineering}

We conducted a structured review of recent work on LLM-assisted safety engineering. The goal was to identify research patterns relevant for medical-device safety analysis, not to perform a full systematic review. We searched Scopus, IEEE Xplore, and Google Scholar, and checked the references of selected papers. Search terms combined LLM-related terms, such as Large Language Models, Generative AI, and ChatGPT, with safety-engineering terms, such as hazard analysis, risk analysis, FMEA, FTA, STPA, HARA, Goal Structuring Notation (GSN), safety cases, assurance cases, and safety documentation. The search was last updated in May 2026.

Papers were included if they applied or discussed LLMs for safety-engineering tasks such as hazard analysis, risk analysis, FMEA, FTA, STPA, HARA, safety cases, safety requirements, or assurance-case review. Papers were excluded if they focused only on general AI safety, clinical diagnosis, cybersecurity without safety analysis, non-LLM safety automation, or general documentation support without a safety-engineering task. The initial search returned more than 200 papers. After title and keyword screening, duplicate removal, abstract screening, and use-case screening, 90 papers published between 2023 and 2026 remained. Due to the limited space in this paper, we have included the list of papers reviewed in an online repository \cite{b_list}.

\begin{table}
\caption{DISTRIBUTION OF REVIEWED PAPERS BY SAFETY TASK AND DOMINANT LLM METHOD}
    \centering
    \renewcommand{\arraystretch}{1.2}
    \begin{tabularx}{\columnwidth}{|p{1.05cm}|X|X|p{0.9cm}|X|X|p{0.55cm}|}\hline
         \textbf{Safety artifact or Task} & \textbf{Discus-sion Paper} & \textbf{Fine tuning LLM}&\textbf{Vector Data-base or RAG} &\textbf{Know-ledge Graph} &\textbf{Prompt Engine-ering} & \textbf{Total}\\\hline
         \textit{FMEA} & 10 & 2 & 8 & 4 & 5 & 29 \\\hline
         \textit{HARA} & 5 & 0 & 4 & 0 & 7 & 16 \\\hline
         \textit{FTA} & 2 & 1 & 0 & 0 & 7 & 10 \\\hline
         \textit{Assurance Cases and Defeater} & 3 & 0 & 2 & 0 & 9 & 14 \\\hline
         \textit{STPA} & 1 & 0 & 0 & 2 & 7 & 10 \\\hline
        \textit{General Safety Analysis} & 5 & 0 & 1 & 2 & 3 & 11 \\ \hline
        \textit{Total} & 26 & 3 & 15 & 8 & 38 & 90 \\ \hline
    \end{tabularx}
    
    \label{tab1}
\end{table}

Table \ref{tab1} classifies the selected papers along two dimensions: the safety task addressed and the dominant LLM method used. The classification is based on the dominant method; most systems also use prompting. Discussion papers propose concepts or discuss opportunities and risks without a full implementation. Fine-tuning denotes papers that adapt an LLM using task-specific training data. Vector database or RAG denotes approaches that retrieve external information from stored documents and provide it to the LLM. Knowledge graph denotes approaches that represent domain and safety knowledge as entities and relations. Prompt engineering denotes approaches where prompting is the main support mechanism and no substantial external retrieval, fine-tuning, or structured knowledge layer is used.

When a paper used several LLM techniques, we assigned the dominant method based on the technique that carried the main safety-engineering contribution. Borderline cases were discussed among the authors and assigned to the category that best reflected the paper’s stated contribution.
The reviewed work shows five main patterns.

First, many studies test feasibility through prompt engineering. They ask whether LLMs can identify hazards, failure modes, unsafe control actions, risk scenarios, safety requirements, or mitigations from a system description \cite{b8, b9, b10, b19, b20, b21, b22, b25}. These studies show that LLMs can produce useful candidate outputs quickly. They also report recurring problems, including missed rare hazards, generic or duplicate entries, implicit assumptions, weak mitigations, and difficulties with structured output formats. Prompt-only use can therefore support early ideation, but it is not sufficient as the sole support method for regulated safety analysis.

Second, retrieval-based grounding is becoming more important. RAG and vector search allow LLMs to use information from prior safety analyses, failure reports, standards, manuals, and technical documents. Such approaches have been studied for performance-level estimation, FMEA generation, STPA support, and safety-requirements generation \cite{b11, b12, b23, b26}. Related medical and radiotherapy studies evaluate LLM support for FMEA-based risk management \cite{b13, b24}. This pattern is important for medical devices because device-specific evidence is often private, technical, versioned, and absent from public model training data.

Third, some studies use structured safety knowledge through knowledge graphs. Knowledge graphs can represent failures, hazards, causes, components, functions, risk controls, and their relations. They have been studied for medical-device risk knowledge extraction and causal analysis in aerospace manufacturing \cite{b15, b16}. FMEA support for human-robot collaboration provides a related example of method-specific safety support \cite{b14}. Compared with plain document retrieval, knowledge graphs can make relations more explicit and may support links from generated safety items back to source evidence.

Fourth, recent work moves from single prompts to multi-step and multi-agent pipelines. Some approaches split safety analysis into smaller tasks, such as hazard extraction, risk-scenario generation, safety-requirement derivation, and test-case generation \cite{b18, b25, b26, b27}. Others use one model to generate candidate artifacts and another model to challenge them. This shift is useful because safety analysis is a process, not a single generation step. However, many such pipelines are still tested on limited examples and do not yet provide enough evidence for repeatability, traceability, or expert usefulness.

Fifth, LLMs are increasingly used as critics, especially in assurance cases. Several papers use LLMs to identify defeaters, meaning doubts or counterarguments to a safety claim \cite{b28, b29, b30, b31}. This line of work is relevant because it treats LLMs not only as generators, but also as support for review and challenge. Similar critique mechanisms could help medical-device experts review candidate hazards, risk controls, residual-risk arguments, and safety-case fragments.

Overall, the literature provides useful starting points but does not yet solve the regulated medical-device problem. Current approaches are often method-specific, prompt-centered, and evaluated on public or small case studies. Only a small part of the literature addresses medical-device safety analysis directly \cite{b13, b15, b24}. The next section derives the resulting gap: LLM support must move from isolated safety-text generation to source-linked safety-knowledge support.

\section{Research Gap: From Generated Text to Source-Linked Safety Knowledge}

The reviewed literature shows that LLMs can support safety-engineering tasks, but current approaches are not yet sufficient for regulated medical-device risk management. The main gap is not simply that LLMs can produce incorrect text. The deeper problem is that most approaches do not provide a controlled process for linking outputs to device evidence, flagging uncertainty, supporting expert review, and updating safety artifacts over the lifecycle.

Table \ref{tab2} summarizes the resulting requirements for a controlled LLM support process. These requirements motivate the research questions and the framework in the next section.

\begin{table}
\caption{RESEARCH GAPS AND REQUIREMENTS FOR LLM-SUPPORTED MEDICAL-DEVICE SAFETY ANALYSIS}
    \centering
    \renewcommand{\arraystretch}{1.2}
    \begin{tabularx}{\columnwidth}{|p{1.4cm}|X|X|}\hline
    
         \textbf{Gap} & \textbf{Relevance for medical devices} & \textbf{Framework Requirement} \\\hline
         
         \textit{Weak Domain grounding} & Device-specific safety knowledge is often private, product-specific, and absent from public model training data. & Controlled retrieval from company and device evidence. \\\hline
         
         \textit{Limited Hazard Coverage} & Severe harm can arise from rare, indirect, or interaction-based scenarios. & Support systematic search for common, rare, and interaction-based hazards. \\\hline
         
         \textit{Limited Traceability} & Safety artifacts must be justified during review, audits, and change analysis. & Link each generated item to source documents, requirements, design elements, risk controls, and evidence. \\\hline
         
         \textit{Unsupported claims and weak scoring} & Plausible but unsupported claims or poor risk scores can affect risk-control decisions. & Flag weak source support, uncertainty, inconsistency, and low-confidence outputs. \\\hline
         
         \textit{Unclear human role} & LLMs must not replace qualified safety engineers. & Require expert review, approval, change tracking, and recorded justification. \\\hline
         
        \textit{Weak evaluation} & Public examples may overlap with training data, and text metrics are insufficient for safety artifacts. & Use non-public or newly built case studies, expert references, and safety-specific metrics. \\\hline
        
        \textit{Limited lifecycle support} & Risk-management files must change with requirements, software, complaints, and post-market data. & Support artifact retrieval, impact analysis, proposed updates, and re-approval. \\\hline
        
    \end{tabularx}
    
    \label{tab2}
\end{table}

A first challenge is \textbf{domain grounding}. Prompt-only use can support early brainstorming, but it is weak for medical-device safety analysis. Critical knowledge is often contained in requirements, software design documents, risk-management files, verification reports, complaints, and post-market reports. Public model training data is unlikely to contain this device- and company-specific evidence. LLM outputs may therefore be plausible but weakly supported by the actual device.

A second challenge is \textbf{hazard coverage}. Medical-device safety analysis must consider common failures as well as rare, indirect, interaction-based, or context-dependent scenarios that may lead to severe harm. LLMs may miss such scenarios or produce duplicate hazards, generic failure descriptions, missing causal chains, or weak risk controls. The goal is not to prove complete hazard identification, but to improve coverage and make remaining uncertainty visible to reviewers.

A third challenge is \textbf{traceability}. Safety artifacts must be linked to requirements, design elements, risk controls, verification evidence, and residual-risk decisions. LLM-generated text without source links is difficult to review, justify during audits, or update when the device changes. Each generated safety item therefore needs an explicit link to the evidence and assumptions that support it.

A fourth challenge is \textbf{hallucination and uncertainty}. LLMs can produce plausible but unsupported or incomplete claims. This is especially critical when they assign severity, occurrence, detectability, risk priority numbers, or residual-risk judgments. Such values influence risk-control decisions and should be treated as candidates for expert review, not as model decisions.

A fifth challenge is the \textbf{human role}. Existing work often includes an expert either at the end of the process or after selected generation steps, but the expert role is not always defined clearly. In medical-device safety analysis, the LLM should assist qualified engineers, not replace them \cite{b22}. Experts must inspect source evidence before accepting generated hazards, causes, risk controls, risk scores, or safety arguments, and their decisions and justifications should be recorded.

A sixth challenge is \textbf{evaluation}. Public case studies are useful for early research, but they create a risk of evaluation-data contamination, where related material may have been present in model training data \cite{b33}. Common natural-language metrics such as BLEU \cite{b32} are also insufficient because they do not assess safety correctness, traceability, risk-control quality, or review value. Evaluation therefore needs expert reference analyses and safety-specific metrics.

A seventh challenge is \textbf{lifecycle support}. Medical-device safety analysis must be updated when requirements, architecture, software, risk controls, verification results, complaints, or post-market data change. Existing LLM studies mostly focus on fixed system descriptions. A useful framework must therefore retrieve affected items, propose updates, and require renewed expert review.

These challenges indicate the need for a framework that treats LLM outputs as candidate safety knowledge items, not as final safety-analysis results.

We use the term source-linked safety item for the central artifact produced, checked, reviewed, and updated in the framework. It is a candidate safety-relevant statement together with the source evidence, assumptions, checks, review status, and change information needed to assess it. For ISO 14971-oriented work, source-linked safety items can represent or support hazards, foreseeable sequences of events, hazardous situations, harms, risk-control measures, verification evidence, and residual-risk decisions.

Each source-linked safety item should contain at least the fields summarized in Table \ref{tab3}. The table is not intended as a fixed data model. It defines a minimal structure that keeps generated safety content reviewable, traceable, and updatable instead of leaving it as untraceable text.

\begin{table}
\caption{CORE FIELDS OF A SOURCE-LINKED SAFETY ITEM}
    \centering
    \renewcommand{\arraystretch}{1.2}
    \begin{tabularx}{\columnwidth}{|p{1.3cm}|X|X|}\hline
    
         \textbf{Field} & \textbf{Purpose} & \textbf{Example} \\\hline
         
         \textit{Item ID} & Stable reference for review and change tracking & HS-1 \\\hline
         
         \textit{Item type} & Hazard, hazardous situation, failure mode, risk control, claim, or review question & Hazardous situation \\\hline
         
         \textit{Device Link} & Function, component, software item, use context, or patient/user interaction & Occlusion detection and alarm function \\\hline
         
         \textit{Source evidence} & Document, section, version, requirement, risk-control record, or artifact link & Software requirement for alarm timing; prior FMEA row on occlusion; verification test for alarm latency; complaint record on delayed alarm response. \\\hline
         
         \textit{Generated statement} & Candidate safety-relevant statement proposed by the LLM & Delayed occlusion alarm may lead to continued under-infusion without timely clinical response. \\\hline
         
        \textit{Rationale} & Causal or method-specific reason why the item was proposed & A changed detection threshold may increase alarm delay under low-flow conditions. \\\hline
        
        \textit{Assumptions} & Conditions that require expert review & Clinical staff rely on the device alarm as the main signal for occlusion response. \\\hline

        \textit{Uncertainty flags} & Weak source support, conflict, duplicate, missing evidence, or model disagreement & Requires source check for low-flow cases; possible duplicate of existing occlusion hazard. \\\hline

        \textit{Reviewer decision} & Accepted, rejected, edited, merged, or split & Edit and merge with existing hazardous situation. \\\hline

        \textit{Reviewer justification} & Short reason for important review decisions & Existing item covers occlusion, but alarm timing change adds a new causal path and affected verification evidence. \\\hline

        \textit{Change history} & Later changes, affected artifacts, and re-approval status & Linked to software requirement change and re-approval of alarm verification test. \\\hline
        
    \end{tabularx}
    
    \label{tab3}
\end{table}

\section{Framework for Evidence-Grounded Safety-Knowledge Support}

The discussed research gap leads to the following research questions. The questions define the requirements for an evidence-grounded LLM support framework for medical-device safety analysis. Table \ref{tab4} maps each question to the framework component that addresses it.

\begin{table}
\caption{RESEARCH QUESTIONS (RQ) AND CORRESPONDING FRAMEWORK COMPONENTS}
    \centering
    \renewcommand{\arraystretch}{1.2}
    \begin{tabularx}{\columnwidth}{|p{0.6cm}|X|p{2cm}|}\hline
    
         \textbf{RQ\#} & \textbf{Research Question} & \textbf{Framework Component} \\\hline
         
         \textit{RQ1} & Which medical-device lifecycle artifacts are needed as evidence for LLM-supported safety analysis? & Evidence Intake \\\hline
         
         \textit{RQ2} & How should device and safety knowledge be represented and retrieved while preserving source links and versions? & Knowledge storage and retrieval \\\hline
         
         \textit{RQ3} & How can LLMs generate method-specific candidate safety items for FMEA, STPA, FTA, assurance cases, and ISO 14971 risk-management work? & Method-specific generation \\\hline
         
         \textit{RQ4} & How can unsupported, inconsistent, duplicate, or low-confidence items be detected before expert review? & Critique and uncertainty checks \\\hline
         
         \textit{RQ5} & How should expert review, approval, rejection, correction, and justification be recorded? & Human review \\\hline
         
        \textit{RQ6} & How can affected safety items be found and updated after design changes, software changes, complaints, or post-market feedback? & Lifecycle update \\\hline
        
        \textit{RQ7} & How can the framework be evaluated fairly using non-public or newly built medical-device cases and expert references? & Evaluation \\\hline

    \end{tabularx}
    
    \label{tab4}
\end{table}

The proposed framework defines a human-in-the-loop support process that addresses the research questions shown in Table \ref{tab4}. It is not a layered software architecture. Rather, it consists of process components connected by artifact and control-flow interfaces. The main artifact passed between the components is the source-linked safety item introduced in Section IV. The framework prepares, links, checks, and updates candidate safety artifacts for expert decision-making; final safety responsibility remains with qualified experts.

\begin{figure*}
  \centering
  \includegraphics[width=0.9\textwidth]{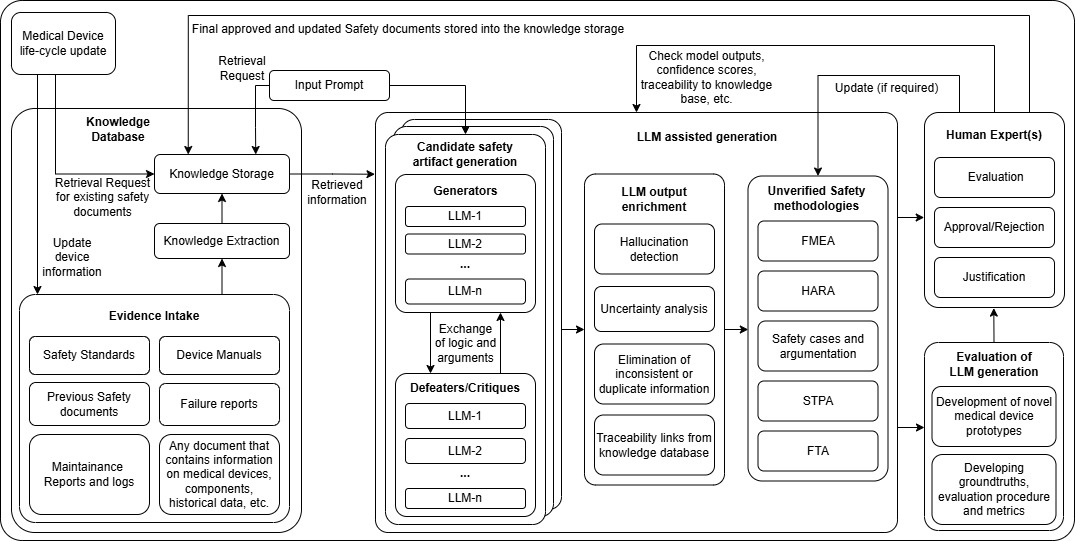}
  \caption{Process components and interfaces of the proposed evidence-grounded LLM support framework.}
  \label{fig1}
\end{figure*}

Figure \ref{fig1} shows the process components and interfaces of the proposed framework. The figure should be read as a support process, not as an automated safety-assessment or compliance-checking pipeline. Evidence and lifecycle artifacts are ingested into a safety-knowledge store. Task and method setup then guide method-specific generation of source-linked safety items. Critique and uncertainty checks add review signals. Qualified experts accept, reject, edit, merge, or split candidate items and record justifications. Lifecycle changes trigger retrieval of affected items and renewed expert review.

\subsection{Evidence intake}

The evidence intake component collects the artifacts needed for safety analysis, such as product descriptions, intended-use statements, requirements, design and software documents, user information, risk-management files, verification evidence, complaints, post-market reports, and relevant standards. For medical devices, these artifacts should be aligned with ISO 14971 risk-management activities and, where applicable, IEC 62304 software lifecycle information.

The intake component should preserve document identity, version, source, date, and artifact type. This is important because later generated safety items must point back to the evidence used during generation and review. A generated risk control based on an outdated requirement should be treated differently from one based on the current approved requirement set.

\subsection{Safety-knowledge storage and retrieval}

The safety-knowledge storage and retrieval component stores and retrieves device and safety knowledge. The collected artifacts may be represented in vector databases, knowledge graphs, structured tables, or hybrid forms. Vector search supports similarity-based retrieval from large text collections. Knowledge graphs make relations between hazards, causes, components, functions, risk controls, and evidence explicit. Structured tables preserve method-specific artifacts such as FMEA rows, risk-control records, and verification links.

The goal is not only to retrieve context for the LLM. The goal is to preserve source links between generated safety items and supporting evidence. Retrieval should therefore return not only text, but also artifact identifiers, versions, sections, and relation types. This makes later review and update analysis possible.

\subsection{Task setup}

The task setup component defines the device or device function under analysis, the system boundary, the intended use, the use environment, relevant users, patient context, and selected safety method. The framework should not assume that one safety method fits all cases. FMEA may support component-level failure analysis. STPA may support unsafe interaction and control analysis. FTA may support top-event reasoning. Assurance cases may support structured argumentation. ISO 14971 provides the risk-management process context that connects these artifacts to risk-control decisions and residual-risk evaluation.

\subsection{Method-specific generation}

The generation component uses the task definition and retrieved evidence to produce source-linked safety items. Depending on the selected method, these items may include hazards, hazardous situations, failure modes, causes, effects, risk controls, safety requirements, verification ideas, safety-case claims, or review questions. Each generated item should include links to the evidence used during generation and should record assumptions that require expert review.

For example, consider an infusion pump software change that modifies occlusion detection and alarm timing. The framework should retrieve related requirements, software items, previous FMEA entries, alarm-related risk controls, verification evidence, complaint data, and prior reviewer decisions. The LLM may then propose candidate updates such as a changed hazardous situation, an affected risk control, or a new verification question. These proposals are not accepted automatically. They become source-linked safety items that must pass critique checks and expert review.

\subsection{Critique and uncertainty checks}

The critique component checks generated items before expert review. This can include duplicate detection, format checks, consistency checks, source-evidence checks, comparison across multiple LLM outputs, and generation of defeaters or review questions. If models disagree, if source support is weak, or if an item relies on an unstated assumption, the framework should flag it for careful review.

These checks cannot prove correctness. They are review aids. Their purpose is to help experts focus on uncertain, weakly supported, inconsistent, or safety-critical items. In medical-device risk management, this is especially important for risk scoring. Severity, occurrence, detectability, and residual-risk judgments should not be accepted merely because they are generated in a plausible form.

\subsection{Expert review and lifecycle updates}

The expert review component is central to the framework. Generated safety items must be reviewed by qualified experts before they are accepted. The reviewer should inspect the source evidence and then accept, reject, edit, merge, or split generated items. The reviewer should also record short justifications for important decisions, especially for rejected hazards, accepted risk controls, residual-risk decisions, and changes to risk scores.

The same structure supports lifecycle updates. When requirements, software, design elements, risk controls, verification results, complaints, or post-market data change, the framework should identify affected source-linked safety items and propose updates. For example, a changed software requirement should trigger retrieval of linked hazards, affected risk controls, related verification evidence, and prior reviewer decisions. These proposed updates must again be reviewed and approved by experts.

The framework therefore shifts LLM use from one-time safety-text generation to lifecycle-oriented safety-knowledge support. Its novelty is not the isolated use of retrieval, knowledge graphs, LLM generation, critique, or expert review. These elements already appear in parts of the literature. The contribution is their integration around a common reviewable artifact: the source-linked safety item. This artifact binds generated safety content to source evidence, method context, assumptions, uncertainty flags, reviewer decisions, and lifecycle history. It therefore gives LLM support a structure that can be reviewed, audited, updated, and evaluated in regulated medical-device development.

\section{Evaluation Strategy}

A useful evaluation must test whether the framework improves safety-engineering support without weakening expert control. We therefore propose an evaluation based on non-public or newly built medical-device case studies, expert reference analyses, and safety-specific metrics.

The case studies should include at least two device types with different safety characteristics. One case should be a software-intensive but non-AI device, such as an infusion-pump function or alarm subsystem. A second case should include an AI-enabled or connected function, where evidence may include model limits, data assumptions, use conditions, post-market feedback, or changes in input distribution. The use of non-public or newly built cases reduces the risk that the LLM has already seen related safety material during training and helps limit evaluation-data contamination \cite{b33}.

Expert reference analyses should be created by qualified safety and domain experts. These references should not be treated as perfect ground truth, but as review baselines. They should include accepted hazards or hazardous situations, causal explanations, risk controls, and verification links. Where experts disagree, the disagreement should be recorded rather than hidden, because such disagreement is common in safety analysis.

The evaluation should compare at least three configurations: prompt-only generation, retrieval-grounded generation, and the full framework with source-linked safety items, critique checks, and expert review. This comparison tests whether the additional process elements add value beyond basic LLM prompting.
We propose the following metrics:

\begin{itemize}
    \item Coverage: share of expert-reference items found by the framework.
    
    \item Correctness: share of generated items judged technically and safety-wise correct by experts.
    
    \item Relevance: share of generated items relevant to the selected device function and safety method.
    
    \item Duplicate rate: share of items that repeat existing items without adding useful detail.
    
    \item Source support: share of accepted items with sufficient source evidence. Source support can be rated by experts as direct support, partial support, no support, or contradiction.
    
    \item Unsupported-claim rate: share of generated items that make claims not supported by the provided evidence.
    
    \item Traceability: share of items linked to requirements, design elements, risk controls, verification evidence, or post-market data.
    
    \item Review effort: expert time needed to accept, reject, edit, merge, or split generated items.
    
    \item Review usefulness: expert judgment of whether the framework helped identify, check, or update safety artifacts.

\end{itemize}

The full framework would be considered useful if, compared with prompt-only and retrieval-grounded baselines, it increases the share of relevant and source-supported safety items, reduces unsupported claims and duplicate items, improves traceability, and reduces expert search and consistency-checking effort. It should not be considered successful if it only increases the number of generated hazards without improving review quality or source support.

The proposed evaluation has several threats to validity. Expert reference analyses may be incomplete or reflect expert-specific judgment. Non-public or newly built case studies reduce training-data overlap, but cannot prove that no related knowledge was present in model training data. Review effort may also depend on reviewer experience, device familiarity, and the quality of the existing risk-management file. We therefore treat the evaluation as evidence of usefulness and review quality, not as proof of complete hazard identification or regulatory sufficiency.

\section{Conclusion}

This paper argued that LLM support for medical-device safety should not be framed as autonomous safety-text generation. Used in that narrow way, LLMs risk adding more unverified text to an already document-heavy process. A stronger role is source-linked safety-knowledge support: connecting device evidence, safety methods, critique checks, expert review, and lifecycle updates around reviewable safety items.

For the medical-device industry, the value hypothesis is specific. Manufacturers need to reuse internal knowledge from requirements, design files, risk files, verification reports, complaints, and post-market data. They also need to keep safety artifacts consistent when devices change. The proposed framework is intended to reduce search, drafting, and consistency-checking effort while keeping qualified experts responsible for final safety decisions. The intended industry benefit is strongest where manufacturers must maintain risk-management files across many device variants, software versions, and post-market feedback loops.

In the MedSafe project \cite{b34}, we will implement and evaluate evidence intake, safety-knowledge retrieval, method-specific generation, critique and uncertainty checks, expert review, and lifecycle updates for selected safety-analysis and risk-management processes. The expected result is not an automatic safety assessor, but a support system for testing whether medical-device safety analysis can become more efficient, reviewable, traceable, and updatable while preserving human responsibility.




\end{document}